\documentclass[10pt, conference, letterpaper]{IEEEtran}

\usepackage{graphicx}
\graphicspath{{../pdf/}{../jpeg/}{../PNG/}}
\usepackage{amsmath}
\DeclareGraphicsExtensions{.pdf,.jpeg,.png}

\usepackage{mathtools}
\usepackage{amsmath,amssymb}
\usepackage{amsfonts}
\usepackage{braket}

\usepackage[tableposition=top]{caption} 
\usepackage{lipsum}
\usepackage{cite}                                                          
\usepackage{stackengine}
\usepackage[font=footnotesize]{caption}

\usepackage{eqparbox}
\usepackage{subcaption}
\usepackage{url}
\usepackage[tableposition=top]{caption} 
\usepackage{lipsum}

\usepackage{array}
\usepackage{mdwmath}
\usepackage{mdwtab}

\usepackage{bibentry}  
\usepackage{xcolor,soul,framed} 
\usepackage{multicol}

\usepackage{epsfig} 
\usepackage{mathptmx} 
\usepackage{times} 

\title{\LARGE \bf
Experimental Challenges of Implementing Quantum Phase Estimation Algorithms on IBM Quantum Computer  
}

\author{Hamed Mohammadbagherpoor$^{1}$, Young-Hyun Oh$^{2}$, Anand Singh$^{2}$, Xianqing Yu$^{2}$, Andy J. Rindos$^{2}$\\  
$^{1}$Department of Electrical and Computer Engineering, \\
North Carolina State University, Raleigh, NC 27606, USA\\
$^{2}$Hybrid Cloud, IBM, Durham, NC, USA
}

\usepackage{lastpage}
\begin{document}

\maketitle

\begin{abstract}
Many researchers have been heavily investigated on quantum phase estimation (QPE) algorithms to find the unknown phase, since QPE is the core building block of the most quantum algorithms such as the Shor's factoring algorithm, quantum sampling algorithms, and finding the eigenvalues of unitary matrices. Kitaev's algorithm and QPE algorithms using inverse Quantum Fourier transform were proposed and widely used by researchers as a key component for their quantum algorithms. In this paper, we explore the experimental challenges of QPE algorithms on Noisy Intermediate-Scale Quantum (NISQ) computers by implementing various QPE algorithms on the state-of-the-art IBM quantum computer. Our experimental results demonstrate that the accuracy of finding the phase using these algorithms are severely constrained by NISQ's physical characteristics such as coherence time and error rates. To mitigate such physical limitations, we propose modified solutions of these algorithms by reducing the number of control gates and phase shift operations. Our experimental results showed that our solutions can significantly increase the accuracy of the finding phase in near-term quantum computers.
\end{abstract}

\section{INTRODUCTION}
Noisy Intermediate-Scale Quantum (NISQ)~\cite{c23} computers provide a framework which can solve particular problems exponentially faster than classical computers by taking advantages of the computational power coming from the use of superposition and entanglement of the qubits~\cite{c12}.  Classical computers works based on a Load-Run-Read cycle in which the input data is loaded into the system, a program runs and then the output of the program is read. However, in quantum computers there is Prepare-Evolve-Measure cycle wherein the quantum states are prepared as the input, manipulate the input states in quantum computers using the operators and then the results are measured. \cite{c21}. Recently, remarkable progress on quantum computers has been achieved by different industries that enable researchers to implement particular classical algorithms using publicly available near-term quantum computers such as IBM QX~\cite{c21} and Rigetti QPU~\cite{c22}. 

Quantum phase estimation (QPE) is the critical building block for various quantum algorithms such as Shor's algorithm for factoring the prime numbers, quantum chemistry to model the molecules and, Grover's algorithm to search~\cite{c1,c2,c3,c4,c5,c6,c11}. However, implementing QPE algorithms on near-term quantum computers are severely constrained by low reliability and high variability of quantum computers' physical characteristics. For example, Shor's algorithm for factoring 15 (i.e., 3*5) on a nuclear magnetic resonance (NMR) computer is presented in \cite{c18} and the number 21 is factored by implementing qubit recycling in a photonic circuit~\cite{c18}. The largest number factored by actual quantum computer is the number 143 which was implemented on a dipolar coupling NMR System by applying adiabatic quantum computation~\cite{c20}. Although near-term quantum computers are currently limited to a small number of qubits, such experimental approaches will be considerably valuable when we can take full advantage of quantum supremacy in near future. 

The main objective of quantum phase estimation is to determine the eigenvalues of a unitary matrix with an unchanged eigenvector. Two main approaches are used to implement quantum phase estimation; The first approach is to extract the phase information by applying the classical post processing computation after utilizing quantum gate operations as known as Kitaev's algorithm~\cite{c7,c8}. Since Kitaev's algorithm requires some classical post processing after performing Hadamard tests, it is necessary to run a minimal number of trials of Hadamard tests to obtain the phase $k_{th}$ bit position with constant success probability. The other approach is to find the phase information using Quantum Fourier transform in which the phase is estimated by applying quantum inverse Fourier Transform~\cite{c9,c10,c13}. An experimental phase estimation based on quantum Fourier transform was implemented on a three-bit nuclear magnetic resonance (NMR) processor~\cite{c16} but it only used to estimate the eigenvalues of one-bit Grover operators. An implementation of phase estimation algorithm on an ion-trapped quantum computer was proposed to find the eigenstates of the system~\cite{c17}. Lloyd et al. have been shown that quantum computers can speed up of some linear algebraic based machine learning algorithms by applying quantum phase estimation technique such as principle component(PCA), support vector machine (SVM), and K-means algorithms~\cite{c14,c15}. However, QFT approach requires a large number of rotation gates for precision digits to obtain more accurate phase information. Without loss of generality, this means that more rotations gates can dramatically increase readout errors from implementation results of QFT algorithms on near-term quantum computers. Thus, it is fundamental to minimize depth and controlled-rotation gates to increase the accuracy of finding the phase on quantum computers. 

In this paper, we investigate on various quantum phase estimation algorithms but, based on our literature review, it is difficult to find experimental results of QPE algorithms from NISQ computers. To address the lack of experimental results, we implement these algorithms on both IBM QASM simulator and the state-of-the-art IBM QX machine and then analyze experimental challenges of implementing quantum algorithm on real quantum computers. The experimental results show that the accuracy of finding the correct phase decreases as the number of qubits and quantum operations increase. To mitigate the problem, we propose modified solutions of these QPE algorithms by minimizing the number of controlled-gate and phase shift operators. Our experimental results demonstrate that our solutions significantly increase the accuracy of finding correct phase. 

This paper is categorized as follows. Section II describes the basic quantum operations and various phase estimation algorithms such as Kitaev's algorithm, iterative algorithm to estimate the phase, and LLoyd algorithm for phase estimation based on inverse Fourier transform (QFT). In section III, the simulation and experimental results for each method are provided and compared. Finally, the conclusion of the paper is summarized in Section IV.  
\begin{figure}[t]
  \begin{center}
  \includegraphics[width=3.5in,height=1.1in]{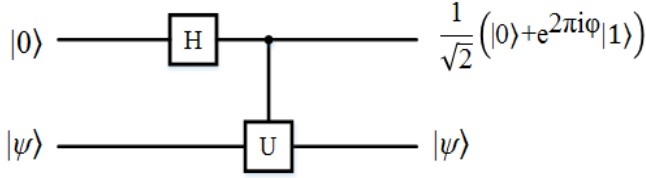}
  \caption{Quantum circuit for transforming the states}
  \label{Ucontrolled}
  \end{center}
\vspace{-1.5em}
\end{figure}

\section{Phase estimation}
Quantum phase estimation of an unitary matrix can be mainly derived using two methods, (1) applying the inverse Quantum Fourier Transform (QFT) to derive the unknown phase information and (2) sequential post processing techniques to calculate the unknown phase. Phase estimation technique is used to estimate the eigenvalues $\ket{\lambda}$ of an unitary matrix $U$ with its known eigenvector $\ket{\psi_\lambda}$~\cite{c12},

\begin{equation}
   U \ket{\psi_\lambda} = \lambda \ket{\psi_\lambda}
\end{equation}
where eigenvalues of the unitary matrix is $\lambda =e^{2\pi i \varphi_m}$. The goal of phase estimation is to find the eigenvalues of the unitary matrix and then applying them to estimate the unknown phase of the unitary operator. Let the phase of the unitary matrix be $\varphi_k = 0.x_1x_2x_3...x_n$ where $n$ is the number of qubits used for phase estimation. The estimated variable $(\hat{\varphi})$ can be expressed as a binary representation,

\begin{equation}
   \hat{\varphi} = \frac{x_1}{2}+\frac{x_2}{2^2}+\frac{x_3}{2^3}+\cdot \cdot\cdot+\frac{x_n}{2^n}
\end{equation}

Fig.\ref{Ucontrolled} shows the quantum circuits are applied to one qubit and eigenstate. The output of the circuits contains the phase described in the top of the Fig.\ref{Ucontrolled} but it is impossible to find the correct phase due to the superposition state on the value. In order to estimate the phase, it requires to apply different QPE techniques to provide the information about the phase of the system. In the next sections, we will explain these techniques along with their simulated and experimentally implemented results.

\begin{figure}[t]
  \begin{center}
  \includegraphics[width=3.5in,height=1.1in]{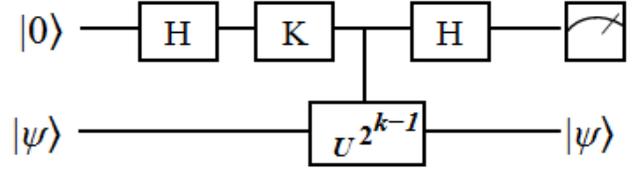}
  \caption{Controlled U circuit}
  \label{Kitaev}
  \end{center}
\vspace{-1.5em}
\end{figure}

\subsection{Kitaev's Algorithm }
Kitaev's algorithm is the first algorithm that was introduced to estimate the phase of an unitary matrix. In this technique a set of Hadamard gates are applied to the input qubits, the outputs of Hadamard gate are performed with Controlled-$U^{2^{k-1}}$ to implement the phase shift operator. Applying controlled-$U$ operator k times transforms the control qubit to $\frac{1}{\sqrt{2}}\big(\ket{0}+e^{-i2\pi\varphi_k 2^{k-1}}\ket{1}\big)$. At each test phase $\varphi_k = 2^{k-1}\varphi$ can be calculated. By doing k times test and measuring the output of each test the set of values ${\varphi, 2\varphi, \cdot\cdot\cdot, 2^{k-1}\varphi}$ can be achieved. These measurements are used to estimate the phase of the unitary matrix. Fig.\ref{Kitaev} shows the circuit to perform phase estimation. As can be seen for the circuit operation K can be used to manipulate the qubit phase and provides more information about the phase of the system. Considering $K=I_2$, the quantum circuit provide the following analysis, 

\begin{equation}\label{Kitaev_eq}
\begin{array}{l}

   \ket{0}\ket{\psi_\lambda} \enspace\>  \xrightarrow{H\bigotimes I} \qquad \frac{1}{\sqrt{2}}(\ket{0}+\ket{1})\ket{\psi_\lambda}\\
   \\
   \qquad\qquad\xrightarrow{C-U_k} \qquad \frac{1}{\sqrt{2}}(\ket{0}+\ket{1})\bigotimes U_k \ket{\psi_\lambda}\\
   \\
   \qquad\qquad    = \frac{1}{\sqrt{2}}(\ket{0}\ket{\psi_\lambda}+e^{2\pi i\varphi_k}\ket{1}\ket{\psi_\lambda})\\
   \\
   \quad \xrightarrow{H\bigotimes I}  \qquad \frac{1}{\sqrt{2}}\frac{\big(\ket{0}+\ket{1}\big)}{\sqrt{2}}\ket{\psi_\lambda} +\frac{e^{2\pi i\varphi_k}}{\sqrt{2}}\frac{\big(\ket{0}-\ket{1}\big)}{\sqrt{2}}\ket{\psi_\lambda}\\
   \\
   \qquad\qquad= \frac{1}{2}\bigg(\big(1+e^{2\pi i\varphi_k}\big)\ket{0} + \big(1-e^{2\pi i\varphi_k}\big)\ket{1}\bigg)\ket{\psi_\lambda} 
    

\end{array}
\end{equation} 

Based on the calculations from Eq.\ref{Kitaev_eq}, the probability of measuring $\ket{0}$ and $\ket{1}$ will be,

\begin{equation}\label{prob_cosine}
   P(0|k) = \frac{1+cos(2\pi\varphi_k)}{2}, \qquad P(1|k) = \frac{1-cos(2\pi\varphi_k)}{2}
\end{equation}
$\varphi_k$ can be obtained with more precise estimated digit by applying more trials. However, based on the data from Eq. \ref{Kitaev_eq} we cannot distinguish between $\varphi_k$ and $-\varphi_k$. Another circuit is required to be considered to provide more information about the phase of unitary matrix and helps to distinguish between $\varphi_k$ and $-\varphi_k$. Combination of the results from $K=I_2$ and $K=S$ helps to find the actual value of the phase. In a case that $K=S$ gate is used and applied in the circuit, the analysis will be,

\begin{equation}
K = S =
\begin{pmatrix}
    1       & 0 \\
    0       & i
\end{pmatrix}
\end{equation}

Quantum circuit provides the following transformation,
\begin{equation}\label{Kitaev_S_part1}
\begin{array}{l}

   \ket{0}\ket{\psi_\lambda} \enspace\> \xrightarrow{H\bigotimes I} \qquad \frac{1}{\sqrt{2}}(\ket{0}+\ket{1})\ket{\psi_\lambda}\\
   \\
    \>\qquad\qquad \xrightarrow{S} \qquad \enspace\enspace\>\frac{1}{\sqrt{2}}(\ket{0}+i\ket{1}) \ket{\psi_\lambda}
    \\
   \qquad\qquad \xrightarrow{C-U_k} \qquad \frac{1}{\sqrt{2}}(\ket{0}+i\ket{1})\bigotimes U_k \ket{\psi_\lambda}\\
   \\
   \qquad\qquad= \frac{1}{\sqrt{2}}(\ket{0}\ket{\psi_\lambda}+i e^{2\pi i\varphi_k}\ket{1}\ket{\psi_\lambda})\\
\end{array}
\end{equation} 
\begin{equation}\label{Kitaev_S_part2}
\begin{array}{l}
   \qquad\quad \xrightarrow{H\bigotimes I}  \qquad \frac{1}{\sqrt{2}}\frac{\big(\ket{0}+i \ket{1}\big)}{\sqrt{2}}\ket{\psi_\lambda} +i \frac{e^{2\pi i\varphi_k}}{\sqrt{2}}\frac{\big(\ket{0}-i\ket{1}\big)}{\sqrt{2}}\ket{\psi_\lambda}\\
   \\
   = \frac{1}{2}\bigg(\big(1+ie^{2\pi i\varphi_k}\big)\ket{0} + \big(1-ie^{2\pi i\varphi_k}\big)\ket{1}\bigg)\ket{\psi_\lambda} 
    \\
   = \frac{1}{2}\bigg(\big(1+e^{2\pi i\varphi_k+\frac{\pi}{2}}\big)\ket{0} + \big(1-ie^{2\pi      i\varphi_k+\frac{\pi}{2}}\big)\ket{1}\bigg)\ket{\psi_\lambda}
    
\end{array}
\end{equation} 
The probabilities in this case will be,

\begin{equation}\label{prob_sine}
   P(0|k) = \frac{1-sin(2\pi\varphi_k)}{2}, \qquad P(1|k) = \frac{1+sin(2\pi\varphi_k)}{2}
\end{equation}

\begin{table}[b]
 \begin{center}
  \includegraphics[width=3.5in]{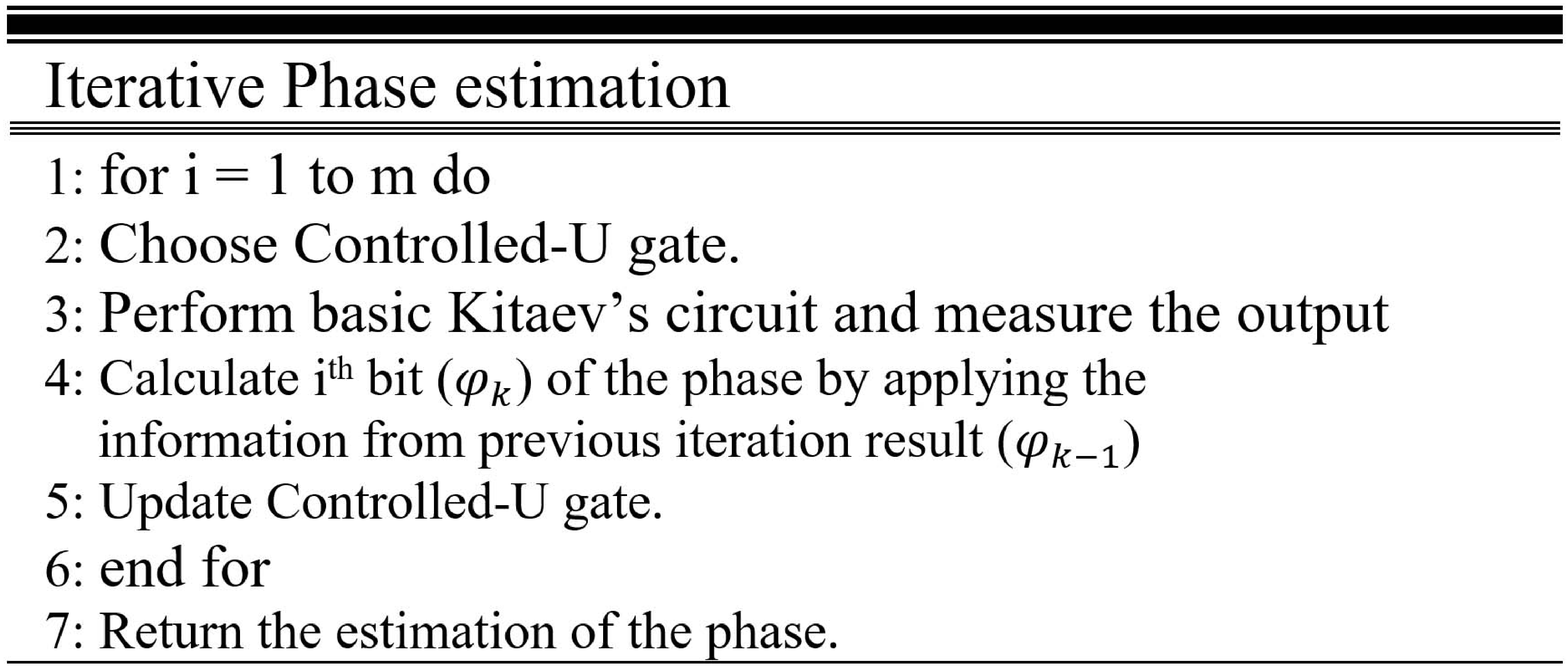}
  \caption{Iterative Quantum phase estimation}
  \label{Iterative_PE}
  \end{center}
\end{table}

Eq.\ref{prob_sine} provides more information about the phase that helps to recover the phase of the unitary matrix. In each test the probabilities of being zero or one in $t$ trials are measured. By using the result from Eq.\ref{prob_cosine} and Eq.\ref{prob_sine} the estimation of $cos(2\pi\varphi_k)$ and $sin(2\pi\varphi_k)$ and as a result hase $(\hat{\varphi})$ can be calculated as, 

\begin{equation}\label{tangant}
   \hat{\varphi_k} = \frac{1}{2\pi}\;tan^{-1} \bigg(\frac{C_k}{S_k}\bigg)
\end{equation}

where $C_k$ and $S_k$ are the estimation of $cos(2\pi\varphi_k)$ and $sin(2\pi\varphi_k)$ respectively. In Kitaev's algorithm post processing calculation is required to estimate the value of the phase. Estimating of the phase within $m$ bits of accuracy requires to increase the number of trials. $O\big(\frac{log(1-\delta)}{\epsilon}\big)$ samples are required to estimate within $\epsilon$ with probability of $1-\delta$.

\begin{figure}[t]
  \begin{center}
  \includegraphics[width=3.5in]{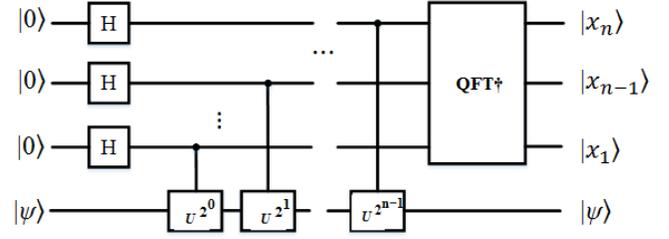}
  \caption{Quantum Phase estimation based on inverse Quantum Fourier Transform (QFT)}
  \label{QFT}
  \end{center}
\end{figure}

\subsection{Iterative Quantum Phase estimation}
In the experimental implementation, increasing the number of gate to estimate the phase with higher accuracy will increase the convergence error and approach to a wrong answer. In this section an iterative technique was proposed to implement the Kiteav's algorithm with higher accuracy and applying two qubits. This sequential approach helps to find the unknown phase of the system with $m$ bits of accuracy. Table \ref{Iterative_PE} shows the the proposed algorithm. As it can be seen, one Hadamard gate was used to perform the superposition and then U-controlled gate is applied to the output of the Hadamard gate then, the output of this stage transform to the measurment state by applying another Hadamard gate. In the next iteration the order of Controlled-U gate is updated and the result from the previous measurement was used and applied to the circuit to estimate the new bit. This technique was repeated $m$ times to estimate the phase with m bits of accuracy. In this method in each iteration the information from the previous iterations were used to estimated the nest bit of the phase 

\subsection{Phase estimation based on inverse QFT }
One of the common method that is used to implement QPE is based on inverse QFT. The general view of this method has been shown in Fig.\ref{QFT}. In this method two stages are required for phase estimation. First stage starts with n-qubits initialized at $\ket{0}$, prepares the state $\ket{\psi}$ and, the second stage uses inverse quantum Fourier transform operation to estimate the binary digits of the phase. The mathematical analysis of the first stage is as,

\begin{equation}\label{Kitaev_eq_sine}
\begin{array}{l}
   \frac{1}{\sqrt{2}}\big(\ket{0}+e^{2\pi i 2^{n-1}\varphi}\ket{1}\big) 
   \frac{1}{\sqrt{2}}\big(\ket{0}+e^{2\pi i 2^{n-2}\varphi}\ket{1}\big)\cdot\cdot\cdot\\
   \frac{1}{\sqrt{2}}\big(\ket{0}+e^{2\pi i \varphi}\ket{1}\big) = \frac{1}{2^{n/2}} \sum_{k=0}^{2^{n-1}}{e^{2\pi i \frac{\varphi k}{2^n}} \ket{k}}
\end{array}
\end{equation}
considering $\varphi = x/2^n$ where $x = \sum_{i=0}^{n-1}{2^i x_i}$ we have,

\begin{equation}\label{QFT_2}
\begin{array}{l}
   \frac{1}{\sqrt{2}}\big(\ket{0}+e^{2\pi i 0.x_n}\ket{1}\big) 
   \frac{1}{\sqrt{2}}\big(\ket{0}+e^{2\pi i 0.x_{n-1}x_n\varphi}\ket{1}\big)\cdot\cdot\cdot\\
   \frac{1}{\sqrt{2}}\big(\ket{0}+e^{2\pi i 0.x_1x_2...x_n}\ket{1}\big) = \frac{1}{2^{n/2}} \sum_{k=0}^{2^{n-1}}{e^{2\pi i \frac{\varphi k}{2^n}} \ket{k}}
\end{array}   
\end{equation}

As can be seen form Fig.\ref{QFT} the outputs from the first stage (phase Kick-back) are the input of inverse QFT. By applying controlled-$U^{2^{n-1}}$there will phase kick back to prepare the states. Also, the output of the first stage is exactly quantum Fourier transform of $\varphi$. By applying the inverse QFT we can recover the unknown phase. In order to analyze this method two different phase estimation circuit with different accuracy have been considered,

\begin{figure}[t]
  \begin{center}
  \includegraphics[width=3.5in]{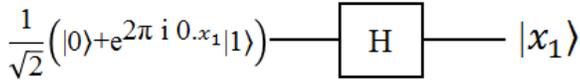}
  \caption{One bit phase estimation Quantum circuit}
  \label{One_digit}
  \end{center}
\vspace{-1.5em}
\end{figure}

\begin{figure}[b]
  \begin{center}
  \includegraphics[width=3.5in]{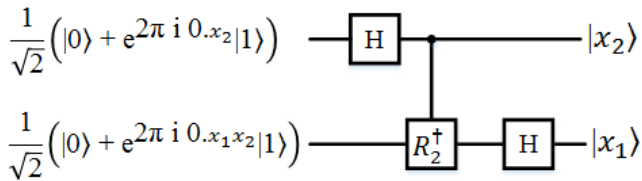}
  \caption{3-Qubits inverse Quantum Fourier Transform (QFT)}
  \label{3_QbitQFT}
  \end{center}
\vspace{-1.5em}
\end{figure}

\textbf{Case1:}
Considering $\varphi$ = $0.x_1$, the circuit shown in Fig. \ref{One_digit} and applying Hadamard gate to the initial state $\ket{0}$ we have:

\begin{equation}\label{Cas1_onedigit}
\begin{array}{l}
   \ket{0}\enspace\>\xrightarrow{H}  \qquad \frac{1}{\sqrt{2}}\big(\ket{0}+ \ket{1}\big)\\

\qquad \xrightarrow{U}  \qquad \frac{1}{\sqrt{2}}\big(\ket{0}+e^{2\pi i\varphi}\ket{1}\big)
    \\
\qquad \xrightarrow{H}     \qquad \frac{1}{2}\big(1+e^{2\pi i\varphi}\big)\ket{0} + \big(1-e^{2\pi i\varphi}\big)\ket{1} 
    \\
\qquad\qquad = \frac{1}{2}\big(1+e^{2\pi i 0.x_1}\big)\ket{0} + \big(1-e^{2\pi i 0.x_1}\big)\ket{1} \\

\end{array}
\end{equation} 

Calculating the probability from equation (\ref{Cas1_onedigit}) we have,

\begin{equation}\label{Probab_case1}
\begin{array}{l}
  P(\ket{0}) = \frac{1+cos(2\pi0.x_1)}{2}, \qquad
  P(\ket{1}) = \frac{1-cos(2\pi0.x_1)}{2}

\end{array}
\end{equation} 

Based on the result from equation (\ref{Probab_case1}), if $x_1=0$, then the probability of $\ket{0}$ is $1$ ($P(\ket{0})=1$) and if $x_1=1$, then the probability of $\ket{1}$ is $1$ ($P(\ket{1})=1$). It means that in a case that phase is considered as one bit only one Hadamard gate is required to extract $x_1$. 


\textbf{Case2:}
Considering $\varphi$ = $0.x_1x_2$, implementing circuit in Fig.\ref{3_QbitQFT} and applying inverse QFT, the unknown phase can be derived. The second digit ($x_2$) can be extracted by applying one Hadarmard gate, the same as case 1. In order to extract the first digit ($x_1$), a controlled-rotation gate $R_2$ is required to remove the impact of the $x_2$ and as a result convert it to case 1 and use one Hadamard gate to estimate $x_1$, as,

\begin{equation}\label{Cas2_twodigit}
\begin{array}{l}

\frac{1}{\sqrt{2}}\big(\ket{0}+e^{2\pi i 0.x_1x_2}\ket{1}\big) \xrightarrow{C-{R^*}_2}   

\\
\qquad \qquad \qquad \qquad \frac{1}{\sqrt{2}}\big(\ket{0}+e^{2\pi i x_1*2^{-1} + x_2*2^{-2} -x_2*2^{-2} }\ket{1}\big) 
\\
\qquad \qquad \qquad \qquad = \frac{1}{\sqrt{2}}\big(\ket{0}+e^{2\pi i x_1*2^{-1}}\ket{1}\big)
\qquad  \xrightarrow{H}  

\\
\frac{1}{\sqrt{2}}\big(1+e^{2\pi i 0.x_1}\big)\ket{0} + \big(1-e^{2\pi i 0.x_1}\big)\ket{1} \\
\end{array}
\end{equation}

\begin{figure}[t]
  \begin{center}
  \includegraphics[width=3.5in]{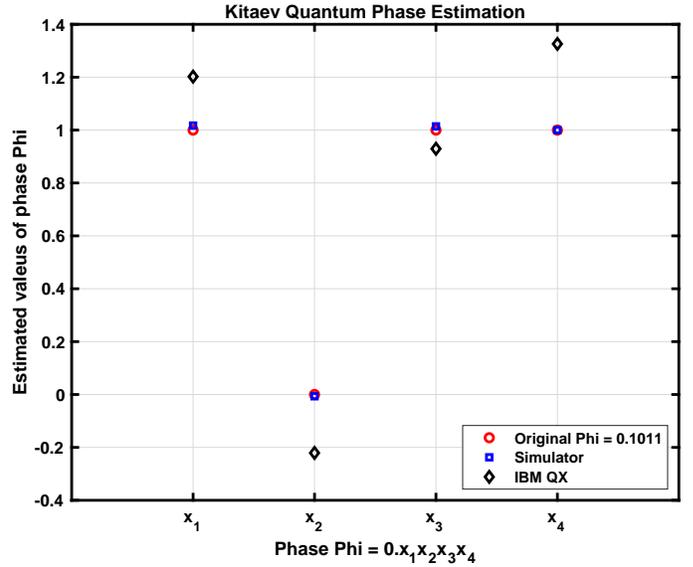}
  \caption{Kitaev Quantum Phase Estimation Algorithm on Simulator and IBM QX}
  \label{kitaev}
  \end{center}
\vspace{-1.5em}
\end{figure}

\begin{figure}[t]
  \begin{center}
  \includegraphics[width=3.5in]{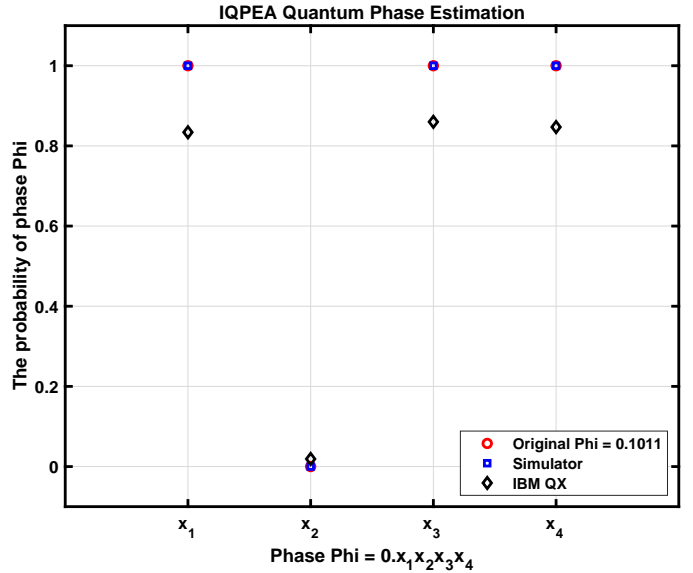}
  \caption{Iterative Quantum Phase Estimation Algorithm on Simulator and IBM QX}
  \label{iqpea}
  \end{center}
\vspace{-1.5em}
\end{figure}

Calculating the probability from equation (\ref{Cas2_twodigit}) we have,

\begin{equation}\label{Probab_case2}
\begin{array}{l}
  P(\ket{0}) = \frac{1+cos(2\pi0.x_1)}{2}, \qquad
  P(\ket{1}) = \frac{1-cos(2\pi0.x_1)}{2}    
    
\end{array} 
\end{equation}

The rotation gate $R_2$ is defined as Eq.(\ref{Rotation_QFT}) where $k=2$.  
\begin{equation}\label{Rotation_QFT}
   R_k = 
\begin{pmatrix}
    1       & 0 \\
    0       & e^{2\pi i/2^k}
\end{pmatrix}
\end{equation}

\section{Simulation and Experiment Results}
In this section, we show nothow to implement various Quantum Phase Estimation (QPE) algorithms on both IBM QASM simulator and the state-of-the-art IBM QX 4 machine (i.e., ibmqx4). Our results show that the accuracy of their experimental results is significantly reduced as the number of qubits increases. For example, single qubit in IBM Q experience has good fidelity on most quantum operations but the fidelity will be quickly decreased as the number of control qubits increases. To mitigate the problem, we implement our modified solutions of these QPE algorithms in order to increase the accuracy of the finding phase. Our solutions not only carefully design quantum gates of QPE algorithms but also take the  advantages of classical computers' capabilities such as storing intermediate results to  feed the values into the next quantum operations. Our simulation results demonstrate that our solutions dramatically increase the accuracy of finding the correct phase. In our experiments, we set the phase is $\varphi = 0.x_1x_2x_3x_4$, where the number of phase bit positions is 4 ($n=4$). We defined $\varphi=1/2 + 1/8 + 1/16$ which represents $\varphi=0.1011$ as a binary value. For each QPE algorithm, we ran the default $1,024$ shots for both simulator and IBM QX machine. 

\begin{figure}[t]
  \begin{center}
  \includegraphics[width=3.5in]{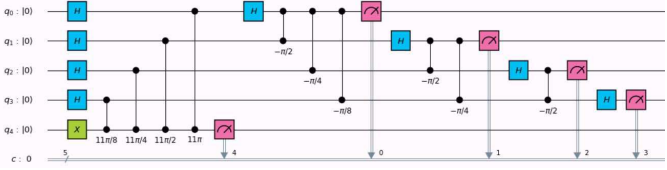}
  \caption{Lloyd QPE Algorithm Gate with 1 ancillary qubit}
  \label{lloyd-orig-gate}
  \end{center}
\vspace{-1.2em}
\end{figure}

\begin{figure}[t]
  \begin{center}
  \includegraphics[width=3.5in]{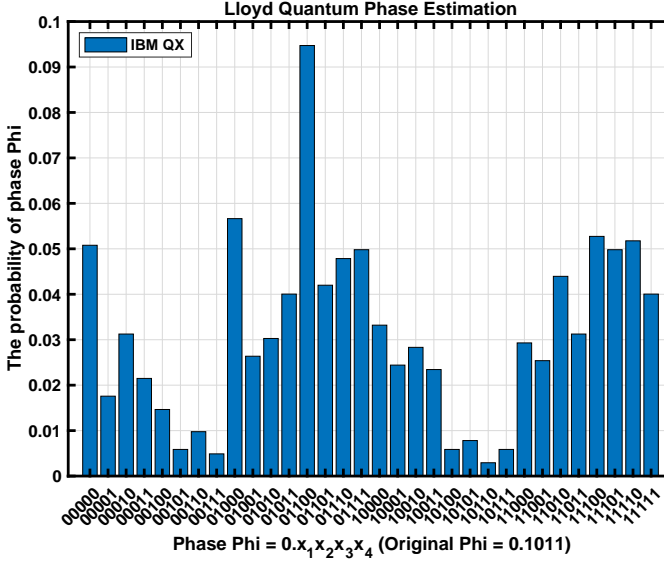}
  \caption{Lloyd QPE Algorithm on IBM QX (ibmqx4)}
  \label{lloyd-orig}
  \end{center}
\vspace{-1.5em}
\end{figure}

First, we implement Kitaev's algorithm to find the phase $\phi$ on both IBM QASM Simulator and IBM QX quantum computer. Fig.\ref{kitaev} shows that the estimated $\hat{\varphi}$ values of simulator results are almost the same as the original $\varphi$ values. The estimated $\varphi$ values from IBM experiment results are slightly different than the original $\varphi$ due to the lack of full error correction capability of quantum computers yet. However, we can estimate the correct binary values of bit positions by converting the estimated $\varphi$ values. Since the noise can be various among different quantum computers, it is critical to find the hardware error rates in order to increase the accuracy of the $\varphi$ estimation in Kitaev's algorithm. The accuracy can be increased by adjusting proper error rates for each quantum computer during the computation process from the estimated $\varphi$ into the binary bit position. 

Second, we implemented iterrative quantum phase estimation algorithm (IQPEA) to find the phase $\varphi$ on both IBM Qasm Simulator and IBM QX quantum computer. Fig.\ref{iqpea} shows that the probability of finding $\varphi$ value from the simulation results are exactly the same as the original $\varphi$. The experiment results are slightly different than the original $\varphi$ but we can estimate the correct binary values the same way as Kitaev's algorithm.


\begin{figure}[t]
  \begin{center}
  \includegraphics[width=3.5in]{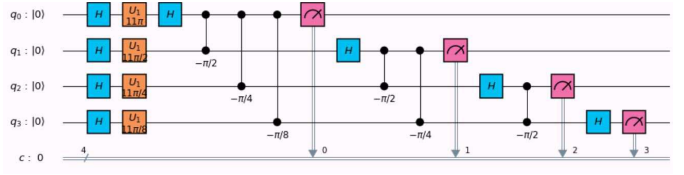}
  \caption{Modified Lloyd QPE Algorithm Gate without 1 ancillary qubit}
  \label{lloyd-gate}
  \end{center}
\vspace{-1.2em}
\end{figure}

\begin{figure}[t]
  \begin{center}
  \includegraphics[width=3.5in]{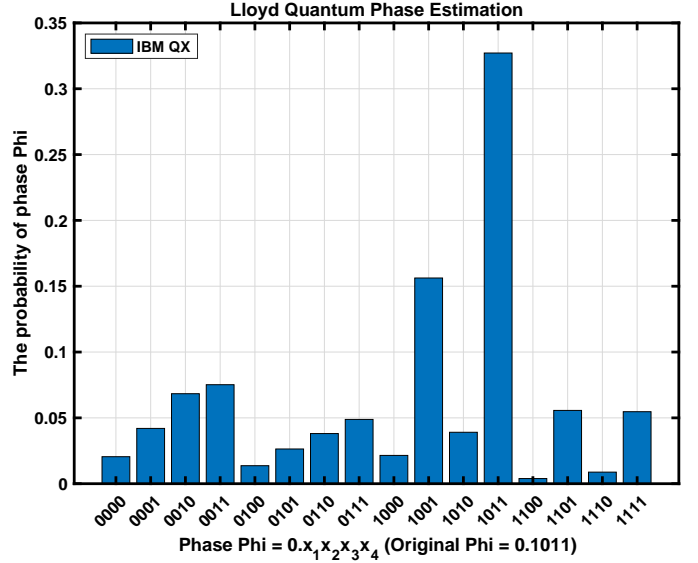}
  \caption{Modified Lloyd QPE Algorithm on IBM QX (ibmqx4)}
  \label{lloyd}
  \end{center}
\vspace{-1.5em}
\end{figure}

Last, we implemented QPE algorithms using inverse Quantum Fourier Transform technique to find the phase $\varphi$ on both IBM Qasm Simulator and IBM QX quantum computer. Fig.\ref{lloyd-orig} only shows the probability of finding $\varphi$ values from IBM QX experiment results because the probability of finding $\varphi$ value from the simulation results are exactly the same as the original $\varphi$. However, the highest probability of the phase $\varphi$ from the experimental results is when the $\varphi$ is $0.01100$ instead of the correct $\varphi=0.1011$. The main reasons for these incorrect results are caused by the lack of error correction capability, short longitudinal, and transversal coherence time for qubits and ancillary qubits respectively. Moreover, as described in Fig.\ref{lloyd-orig-gate}, the number of controlled phase rotation gates on qubits can increase the readout errors. To solve this problem and increase the accuracy of experimental results, we remove the ancillary control qubit and replace the unnecessary controlled-rotation gates with unitary rotation gates for each qubit as described in Fig.\ref{lloyd-gate}. Our experimental results Fig.\ref{lloyd} shows that our solution can find the correct phase $\varphi$ and even the probability (i.e., 0.335$\%$) is completely distinguished from other estimated $\varphi$ values.

\section{CONCLUSIONS}
 This paper investigates on the existing Quantum Phase Estimation (QPE) algorithms and explains how to implement these algorithms on the state-of-the-art IBM quantum computer. We discover the challenges of implementing QPE algorithms on real quantum processor and propose modified solution of these algorithms by minimizing the number of controlled-rotation gates and by utilizing the classical computer's capabilities. Our experimental results can guide researchers to consider these challenges when they implement their quantum algorithms on Noisy Intermediate-Scale Quantum (NISQ) computers before we can fully take the advantages of quatum supremacy in near future.

\addtolength{\textheight}{-12cm}   



\end{document}